\documentclass{elsart}

\begin{document}

\begin{frontmatter}

\title{Projective Dynamics Analysis of Magnetization Reversal}

\author[BROWN]{Gregory Brown\corauthref{cor}},
\corauth[cor]{Corresponding author. FAX: +1-850-644-0098}
\ead{Email: browngrg@csit.fsu.edu.}
\author[NOVOTNY]{M.~A. Novotny},
\author[RIKVOLD]{Per Arne Rikvold}

\address[BROWN]{
         Center for Computational Sciences,
         Oak Ridge National Lab, \\
         Oak Ridge, TN 37831-6114, USA\\
         School of Computational Science and Information Technology, \\
         Florida State University, Tallahassee, FL 32306-4120, USA}

\address[NOVOTNY]{
         Department of Physics and Astronomy and \\
         ERC Center for Computational Sciences,\\
         Mississippi State University, MS 39762, USA}

\address[RIKVOLD]{
        Department of Physics, CSIT, and MARTECH,\\ 
        Florida State University, Tallahassee, FL 32303-4350, USA}

\begin{abstract}
The computational Projective Dynamics method is used to analyze
simulations of magnetization reversal in nanoscale magnetic pillars.
It is shown that this method can be used to determine the
magnetizations corresponding to the metastable minimum and saddle
point in the free energy, and the free-energy barrier associated with
those points. For the nanopillars studied here, entropy is found to
provide a significant contribution to the free-energy barrier which
determines the reversal time scale.
\end{abstract}


\end{frontmatter}

Understanding magnetization reversal is important for technological
applications where the magnetic orientation of nanoscale regions must
be quickly assigned. For information storage, two configurations of
the magnetization are used to encode the state of a bit of
data. During the assignment process, strong fields are applied to
create a free-energy minimum for only one of these configurations, and
the bit almost certainly assumes the equilibrium configuration. For
weaker fields, the two configurations correspond to local free-energy
minima separated by a free-energy maximum. The minimum corresponding
to a magnetization parallel to the applied field is truly stable,
while the antiparallel minimum is higher in energy and therefore
metastable. In this situation the configuration depends not only on
the relative weights of the minima, but for device and human time
scales, it also depends on the history of the configuration. At some
point along the most probable path between the minima there must occur
a free-energy maximum in the form of a saddle point. The free-energy
difference between the maximum and either minimum, often called a
free-energy barrier, determines the time scale for transitions of the
configuration from that minimum to the other.

To a first approximation these barriers and the curvature of the free
energy near the extrema control the nonequilibrium dynamics of a
system. Thus locating free-energy extrema is essential for developing a
detailed understanding of the dynamics of metastable states, and that
understanding is important for such things as maximizing data
integrity and enabling the technology of hybrid recording
\cite{Ruigrok}, which uses lower-than-coercive applied fields to
assign magnetic orientations in high-coercivity magnetic materials.
While the free-energy minima can be easily located, the saddle point
has proven much harder to measure \cite{E}. Here we present results
using the Projective Dynamics method \cite{KOLESIK,NOVOTNY,MITCHELL}
to probe the magnetization reversal of high-aspect-ratio nanoscale
model magnets.

It has already been shown that the Projective Dynamics method can be
used to locate the saddle point \cite{NOVOTNY,UGA02,MMM02}. The method
involves projecting the original description of the dynamics in terms
of a large number of variables onto a stochastic description in terms
of one variable. For example, the dynamics of the thousands of
individual spins in a nanoscale pillar can be projected onto the
stochastic dynamics of the total magnetization along the long axis of
the nanomagnet, $M_z$.  The transition rates between values of $M_z$
are measured by the probabilities $P_{\rm grow}$, which correspond to
an increase in the volume of stable magnetization, and $P_{\rm
shrink}$, which correspond to a decrease in the volume. For values of
$M_z$ with $P_{\rm grow}$$>$$P_{\rm shrink}$, on average the volume of
the stable magnetization grows. This corresponds to a negative local
slope for the free energy. Likewise, $P_{\rm grow}$$<$$P_{\rm shrink}$
corresponds to a positive local slope. Values of $M_z$ for which
$P_{\rm grow}$$=$$P_{\rm shrink}$ correspond to zero slope, {\em i.e.}
extrema of the free energy. Crossings of $P_{\rm grow}$ and $P_{\rm
shrink}$, then, can be used to determine the locations of the extrema,
including the saddle point.

Here we present results for Projective Dynamics applied to
micromagnetic simulations of magnetization reversal in a chain of 17
spins, ${\vec S}_i$, with the chain aligned along the $z$-axis
\cite{BandB}. Using an extended Heisenberg model, the internal energy
of the system is given by
\begin{equation}
E = -J \sum_{i=1}^{16} {\hat S}_i \cdot {\hat S}_{i+1}
    -\frac{D}{2} \sum_{i} \sum_{j \ne i} 
         \frac{ \left[ 3 {\hat z} \left( {\hat z} \cdot {\hat S}_j \right) 
                - {\hat S}_j \right] \cdot {\hat S}_i 
              }
              { |j-i|^3 }
    + B \sum_{i} S_{i,z}
\;,
\end{equation}
where $J$ is the exchange energy, $B$ is the strength of the external
field oriented parallel to $-{\hat z}$, and $D$ is the strength of the
dipole-dipole interactions in energy units, see Eqs.~(17) and (18) of
Ref.~\cite{BrownOpus}. We choose parameters consistent with previous
studies of iron nanopillars modeled as a one-dimensional chain of
spins \cite{MMM02,BandB,BrownOpus,WIRTH}: $J$$=$$1.6 \times 10^{-12}$~erg and
$D$$=$$4.1\times 10^{-12}$~erg.  The dynamics consist of each spin
precessing around a local field, i.e. the Landau-Lifshitz-Gilbert
(LLG) equation \cite{BROWN63,AHARONI}
\begin{equation}
\frac{ {\mathrm d} {{\hat S}_i}}
     { {\mathrm d} {t} }
 =
   \frac{ \gamma_0 }
        { 1+\alpha^2 }
   {{\hat S}_i}
 \times
 \left[
   {{\vec H}_i}
  -{\alpha} {{\hat S}_i} \times {{\vec H}_i}
 \right]
\;,
\end{equation}
where the scaled electron gyromagnetic ratio $\gamma_0 =
9.26\times10^{21}$ Hz/erg for this system, and the local field is
given by the functional derivative ${\vec H}_i = - \delta E / \delta
{\vec S}_i$. The phenomenological damping parameter $\alpha=0.1$ was
chosen to give underdamped dynamics, and an Euler integration scheme
was used \cite{FENG}.

Dipole-dipole interactions make head-to-tail alignment of the spins
along the chain favorable and provide a strong uniaxial anisotropy for
this system. For each simulated reversal, the nanopillar was allowed
to come to equilibrium in a field $B_0$$=$$1.9 \times 10^{-12}$ erg
($1000$ Oe). Then the magnitude of $B$ was quickly decreased to a
value of $-B_0$ to create a metastable configuration of the
magnetization.  For each integration step, data for $P_{\rm grow}$,
$P_{\rm shrink}$, and the energy were binned on $M_z$. Results for
this system have been presented previously \cite{MMM02}. Here we
present improved results obtained from a systematic choice in bin
width and from including lower temperatures.

Projective dynamics is most easily analyzed for discrete systems, such
as the Ising model with spin-flip dynamics. Then $M_z$ changes by
$-S$, $0$, or $+S$ at each step, and the stochastic dynamics are easy
to analyze in terms of jumps along a well-defined chain of states.
When $M_z$ is a continuous variable, binning is used to create the
chain of states. However, there are competing limitations that
determine the bin size. One is the assumption of complete mixing
within a bin, which is required for the transitions between bins to be
a Markov process. If the bins are too large, then complete mixing
within the bins will not be present and there will be memory effects.
Another limiting effect is gathering enough statistics in each bin to
measure the probabilities $P_{\rm grow}$ and $P_{\rm shrink}.$ If the
bins are too small, there will be inadequate statistics related to these
measured probabilities.  Furthermore, we would like to have the
projected Markov matrix be tridiagonal, which allows calculation of
residence times using Eq.(3), introduced below, rather than inverting
a large matrix.  Tridiagonality of the Markov matrix requires that
jumps from anywhere in the bin can go only into the bin with the next
largest or smallest $z$-component of the magnetization.

\begin{figure}[tb]
\vskip 4 in
\includegraphics{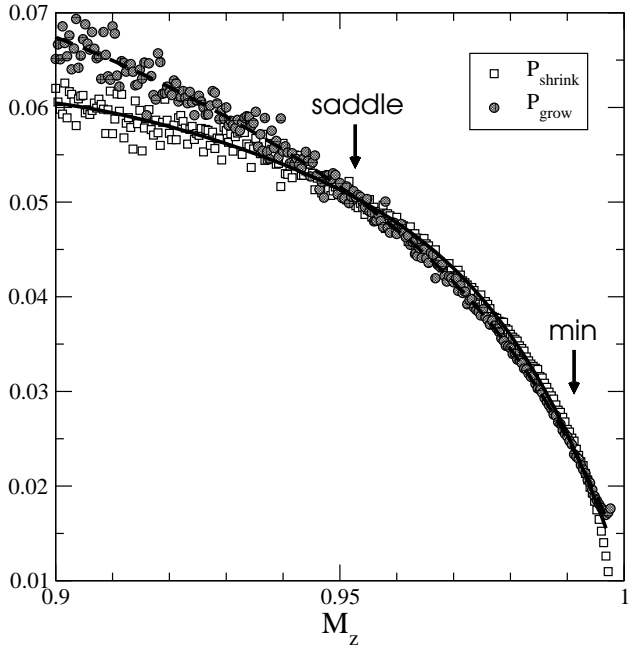}
\caption[]{Probability of shrinking, $P_{\rm shrink}$, (squares) and
growing, $P_{\rm grow}$, (circles) for $T$$=$$5$~K. The solid curves
are fifth-order polynomial fits to the data. The locations of the
metastable minimum and saddle point, as indicated by the arrows, are
determined from the intersections of the polynomials.}
\end{figure}

\begin{figure}[tb]
\vskip 4 in
\includegraphics{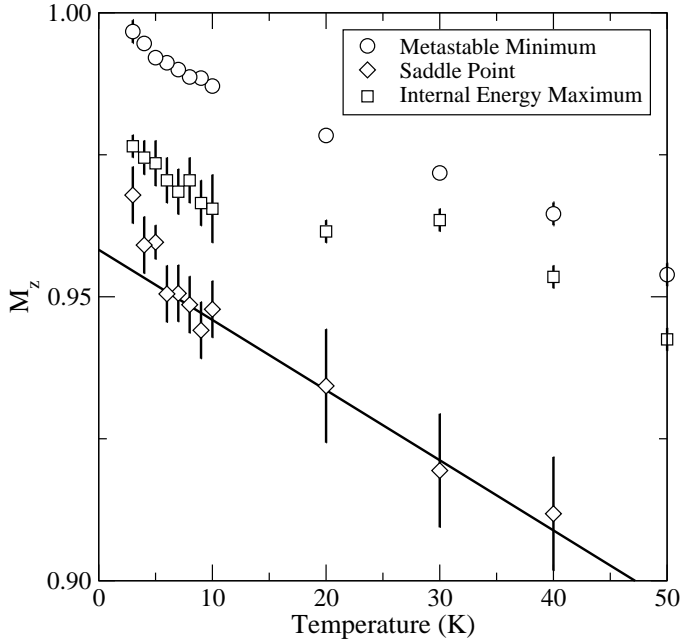}
\caption[]{$M_z$ of metastable minimum (circles), internal energy
maximum (squares), and saddle point (diamonds) vs temperature. The
separation between the saddle point and the internal-energy maximum
implies large entropic effects in magnetization reversal. The solid line
is a least-squares fit to the saddle-point data. The metastable
minimum is identically unity at $T$$=$$0$~K.}
\end{figure}

\begin{figure}[tb]
\vskip 4 in
\includegraphics{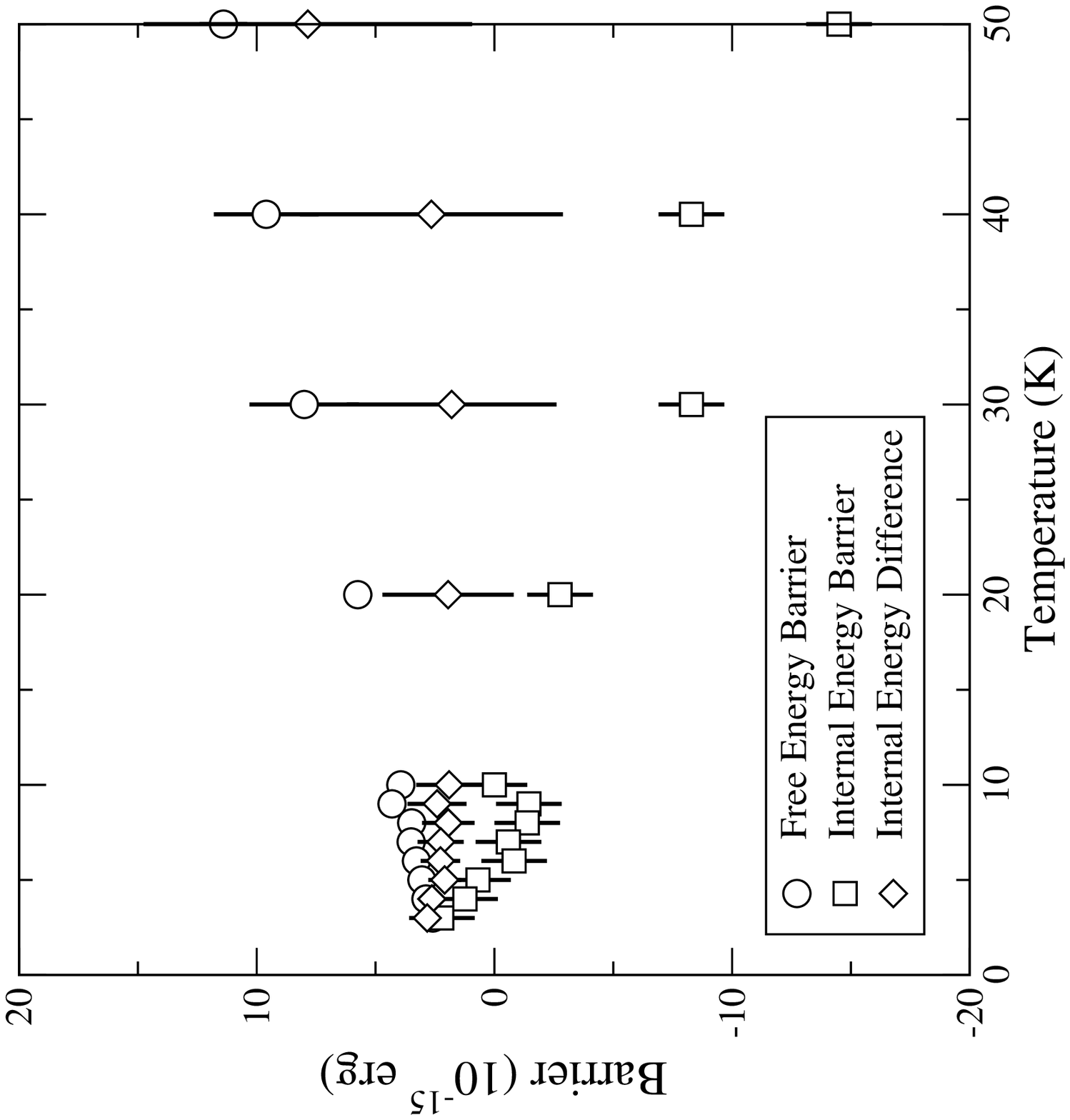}
\caption[]{The free-energy barrier (circles) and internal-energy
barrier (squares) vs temperature. The difference between the two
barriers is due to entropic contributions and vanished as
$T$$\rightarrow$$0$. The difference between the maximum internal
energy and the internal energy at the metastable minimum (diamonds) is
also shown.}
\end{figure}

The results presented in this work were obtained in the following
manner.  Data were collected for on the order of $10^2$ switches for
each temperature with a bin of width $b$$=$$5 \times 10^{-5}.$ The
data were then analyzed for bin widths of $b$, $2b$, $3b$, etc. until
jumps only occurred between neighboring bins. Using this method, the
optimal bin width was found to be $5b$ at $T$$=$$3$~K and $25b$ at
$T$$=$$50$~K. In contrast, in Ref.~\cite{MMM02} bins of width $100b$
where used. Both probabilities, $P_{\rm grow}$ (circles) and $P_{\rm
shrink}$ (squares), for $T$$=$$5$~K, are shown in Fig.~1. Also shown
are two fifth-order polynomials fit to these probabilities.
Throughout this work, fifth-order polynomial fits have been used to
determine the intersection of $P_{\rm grow}$ and $P_{\rm shrink}$, and
we find that the order of polynomial used affects the estimated
location by less than $1\%$ for the saddle point. The crossings for
$T$$=$$5$~K are labeled with arrows in Fig.~1, with the crossing near
$M_z$$=$$0.995$ the metastable minimum and that near $M_z$$=$$0.95$
the saddle point.

The values of $M_z$ for the metastable minimum (circles) and the
saddle point (diamonds) for all temperatures $T$ are shown in
Fig.~2. Similar results have been shown previously \cite{UGA02,MMM02},
but the results shown here are quantitatively different because of the
smaller bins used. The changes are as large as $10\%$ and are
strongest for the saddle point. The linear dependence of the $M_z$ of
the saddle point on $T$ noted previously \cite{UGA02,MMM02} is present
in the new data as well, but only for $T$$>$$5$~K. The solid line is a
least-squares fit, and its intercept is $M_z$$=$$0.958$. The deviation
from linearity at low temperature is expected to the extent that
extrapolation of the linear fit to $T$$=$$0$~K is not consistent with
the internal-energy maximum, described below. This deviation was not
seen in previous studies since such low temperatures are considered
here for the first time.

The free energy, $F$$=$$E-TS$, has contributions from the internal
energy and the entropy, $S$. When the temperature is exactly zero,
there is no entropic contribution to the free energy, and the
saddle point corresponds to the maximum in the internal energy.  The
results for the maximum of the average internal energy, $\langle E
\rangle$, binned on $M_z$ are shown (squares) in Fig.~2. The $M_z$ for
the saddle point and the internal-energy maximum converge as the
temperature decreases, as expected.  The separation between $M_z$ at
the internal-energy maximum and the saddle point at fixed $T$ is a
measure of the importance of entropic contributions to the
magnetization reversal process. The separation in Fig.~2 indicates
that entropy is quite important in these model nanopillars.

The internal-energy ``barrier'', the difference between $\langle E
\rangle$ at the saddle point and the metastable minimum, is shown vs
temperature $T$ in Fig.~3 as squares. Note that for all but the lowest
temperatures the difference is negative, indicating that the $\langle
E \rangle$ for the saddle point is {\em lower} than that of the
metastable minimum. It is indeed possible for a system to be
metastable in this situation, because it is a {\em free-energy}
barrier that is actually required. The free-energy barrier can be
estimated using Projective Dynamics from the time spent in each bin
$i$, $h(i)$, which is proportional to the Boltzmann factor
$\exp(-F(i)/k_BT).$ For a tridiagonal Markov matrix, the residence
time $h(i)$ can be found from the growing and shrinking probabilities
by \cite{NOVOTNY,MITCHELL,KOLESIK}
\begin{equation}
h(i) = \left[1+P_{\rm shrink}(i-1)h(i-1)\right]/P_{\rm grow}(i) \;,
\end{equation}
with $h(1)=1/P_{\rm grow}(1),$ where we take the first bin to be the
stable state and iterate Eq.~(3) towards the metastable state. (Note
that the cited references iterate from the metastable to stable
state.)  free-energy barrier is $\Delta F$$=$$k_BT{\rm ln}(h(i_{\rm
min})/h(i_{\rm saddle}))$, where $i_{\rm min}$ is the bin containing
the metastable minimum and $i_{\rm saddle}$ contains the saddle
point. The results for the free-energy barrier are shown (circles) in
Fig.~3. The free-energy barrier is always positive and approaches the
internal-energy barrier as $T$$\rightarrow$$0$, as expected. However,
it is clear from the internal-energy barrier being negative for most
temperatures and from the temperature dependence of $\Delta F$, that
the entropy contributes significantly to the metastability. This
indicates that the configuration of $\{S_i\}$ is relatively restricted
at the saddle-point, and that this restriction is more important than
the combined dipole-dipole and exchange energies at all but the
smallest temperatures. The difference between the maximum $\langle E
\rangle$ and the value at the metastable minimum is also shown in
Fig.~3 as diamonds. This difference might easily be mistaken for the
barrier to magnetization reversal, but here it is much less sensitive
to the temperature than the measured free-energy barrier.

In this paper, the usefulness of Projective Dynamics for locating the
metastable minimum and saddle point in magnetization reversal has been
demonstrated. Once these points are known, it is also possible to
directly find the free-energy barrier of the reversal process, also
using Projective Dynamics, or the internal-energy barrier, using some
other technique to find the internal energy. One of the truly exciting
aspects of Projective Dynamics is that it is not limited to analyzing
simulations. Since only data on the time-dependence of the slow
variable is needed, the analysis can also be applied to experimental
data.

This work was supported by the LDRD Program of ORNL, managed by
UT-Battelle, LLC (U.S. DOE Contract No. DE-AC05-00OR22725),
by the Computational Material Science Network of BES-DMSE,
by the U.S. NSF (Grant 0120310), and by Florida State University.

\end{document}